\documentclass[letterpaper]{article} 
\usepackage{aaai17}  
\usepackage{times}  
\usepackage{helvet}  
\usepackage{courier}  
\usepackage{url}  
\usepackage{graphicx}  
\usepackage{tabularx}
\usepackage{todonotes}
\usepackage{booktabs}
\usepackage{array}
\newcolumntype{L}[1]{>{\raggedright\let\newline\\\arraybackslash\hspace{0pt}}m{#1}}
\newcolumntype{C}[1]{>{\centering\let\newline\\\arraybackslash\hspace{0pt}}m{#1}}
\newcolumntype{R}[1]{>{\raggedleft\let\newline\\\arraybackslash\hspace{0pt}}m{#1}}
\newcommand{\citeay}[1]{\citeauthor{#1}~\shortcite{#1}}
\copyrighttext{Copyright \copyright 2017\\The copyright remains with the authors.}

\title{From Task Classification Towards Similarity Measures for Recommendation in Crowdsourcing Systems}
\author{Steffen Schnitzer \and Svenja Neitzel \and Christoph Rensing\\
Multimedia Communications Lab - Technische Universit\"{a}t Darmstadt, Germany\\
   {steffen.schnitzer@kom.tu-darmstadt.de}\\
   {svenja.neitzel@kom.tu-darmstadt.de}\\
   {christoph.rensing@kom.tu-darmstadt.de}\\
}
\begin{document}
\maketitle
\begin{abstract}
Task selection in micro-task markets can be supported by recommender systems to
help individuals to find appropriate tasks. Previous work showed that for the
selection process of a micro-task the semantic aspects, such as the
\textit{required action} and the \textit{comprehensibility}, are rated more
important than factual aspects, such as the payment or the required completion
time. This work gives a foundation to create such similarity measures.
Therefore, we show that an automatic classification based on task descriptions
is possible. Additionally, we propose similarity measures to cluster micro-tasks
according to semantic aspects.
\end{abstract}
\section{Introduction and Related Work}\label{sec:intro}
A preceding user study~\cite{schnitzer2015} shows that the similarity of tasks
is an important factor for workers when selecting tasks in crowdsourcing
platforms. Another preceding study~\cite{schnitzer2016} identified the most
important similarity aspects for workers. The semantic aspects, in contrast to
the factual aspects, were found to be the five most highly rated similarity
aspects,  with \textit{required action} and \textit{comprehensibility} coming
first and second.\\
This work provides a foundation to leverage such semantic aspects for
recommending tasks in crowdsourcing platforms using similarity measures based on
task descriptions. To show, that task descriptions in micro-task markets are
diverse and informative enough to support a recommendation, our first approach
classifies tasks into predefined categories.
Therefore, an automatic classification of tasks is implemented and evaluated on
a dataset of 1466 micro-tasks retrieved from the platform \textit{Microworkers}.
We compare different classification approaches by evaluating different feature
sets and their combinations as well as several classification algorithms. This
allows us to conclude, that the employed methods are capable of classifying
micro-tasks into logical categories. On this basis, we conclude that similarity
measures for the identified semantic similarity aspects can be created from task
descriptions. Therefore, we propose a first idea for creating similarity
measures based on task descriptions considering the semantic aspects
\textit{required action} and \textit{comprehensibility}.\par
One approach to task recommendation has been proposed by \citeay{yuen2012},
considering task properties, worker performance and history of the worker's
completed tasks. The two methods proposed and compared by \citeay{ambati2011}
use a classification as well as an approach based on semantic similarities.
\citeay{mavridis_using_2016} apply a taxonomy based skill modeling approach to
optimize task assignment quality. Within our classification approach, the
textual information from the micro-tasks is used to classify them into the
categories provided by the platform. \citeay{arora_good_2015} present an
approach for classifying questions posted to a Q{\&}A platform. In a very similar domain
to micro-tasks, \citeay{schnitzer_combining_2014} and \citeay{schmidt_text_2016}
use a \textit{tf-idf} based approach and an ensemble classifier in order to
classify job offers.
\section{Task Classification}\label{sec:dataset}
\subsubsection{Dataset and Preprocessing}
The dataset of 1466 micro-tasks was gathered between October and December 2015
from the micro-task market platform \textit{Microworkers}. For each task we
extract the ID, title, description, proof, category, employer, payment, time to
finish, time to rate, no. of jobs available/done, success rate and countries the
task is available in. A number of common preprocessing steps are applied to the
textual task attributes and some meta information about the original text given
by the HTML structure is extracted and stored as additional attributes.
\subsubsection{Classification of Micro-Tasks}\label{sec:classification}
For classification, the machine learning tool Weka is used. To identify the most accurate setup for classification, four different feature sets (see Table~\ref{table:4_features}) are extracted and six different classifiers are trained on every combination of the feature sets.
\begin{table}
	\begin{center}
		\caption[Feature sets.]{Feature sets.}
		\label{table:4_features}
		\begin{tabular}{lL{5.5cm}}
			\toprule
			\textbf{Feature Set}	& \textbf{Features}																			\\ \midrule
			factual					& payment, time to rate, time to finish, positions, payment per minute, employer, countries	\\ \midrule
			content					& n-grams																					\\ \midrule
			structural				& word count, no. of bullet points, avg. words per sentence, avg. commas per sentence, 
									avg. chars per word, avg. paragraph length,	avg. line length, readability 
									(Gunning Fog Index~\cite{gunning1952}), lexical diversity~\cite{dickinson2015} 				\\ \midrule
			semantic				& URL hosts, named entities, sentiment														\\
			\bottomrule
		\end{tabular}
	\end{center}
\end{table}
The Weka implementations of six different classifiers are used to evaluate the performances of Naive Bayes, Random Forest, K Nearest Neighbors (\textit{IBk}), Support Vector Machine (\textit{SMO}), a rule based classifier (\textit{JRip}) and a decision tree (\textit{J48}).
\subsubsection{Evaluation}\label{sec:evaluation}
A 10-fold stratified cross-validation is executed on the dataset for each
classifier using each feature set. The results obtained for the three best
performing classifiers (JRip, SMO and Random Forest) in terms of weighted
average F1-score are given in Table~\ref{table:5_results}. The content feature
set using a tf-idf approach achieves the best results over all classifiers. The
SMO classifier obtains the highest F1-score of 0.94 using the content feature
set. However, it is outperformed by the two other classifiers when using the
factual or structural feature set. Random Forest turns out to be the most stable
classifier across the four feature sets.\\
\begin{table}
	\begin{center}
		\caption[F1-scores for different classifiers and feature sets.]{F1-scores for different classifiers and feature sets.}
		\label{table:5_results}
		\begin{tabular}{lrrr}
			\toprule
			\textbf{Feature Set}  & \multicolumn{1}{r}{\textbf{Random Forest}} & \multicolumn{1}{r}{\textbf{JRip}} & \multicolumn{1}{r}{\textbf{SMO}}\\
			\midrule
			factual					& 0.86			& 0.82			& 0.73			\\
			structural				& 0.81			& 0.74			& 0.54			\\
			semantic				& 0.83			& 0.75			& 0.84			\\
			content (tf-idf)		& \textbf{0.92}	& \textbf{0.92}	& \textbf{0.94}	\\
			\bottomrule
		\end{tabular}
	\end{center}
\end{table}
This evaluation shows in general, that it is possible to reproduce the task
categories and that a classification of micro-tasks is feasible. Content
features were shown to be the best performing feature set, while the SMO
classifier provided the best results among the classifiers. A per class
evaluation showed further, that all classes with at least 10 examples can be
classified with an F1-score above 0.7.
\section{Similarity Measures for Micro-Tasks}\label{sec:similarities}
Task similarities based on the semantic aspects \textit{required action} and
\textit{comprehensibility}, that were found to be relevant in the preceding
study \cite{schnitzer2016}, cannot be produced by a classification approach. The
classification considers binary category membership, while similarities rely on
continuous measures. However, the insights about the applicability of certain
features for the classification task can be used and extended to propose an
approach for calculating task similarities based on these aspects. As there is
no labeled data and no predefined classes for \textit{required action} and
\textit{comprehensibility} for micro-tasks, an unsupervised approach is
necessary. In the following we propose certain features for the similarity
measures that are specific for each of the semantic aspects.
\subsubsection{Features for \textit{required action}}
To measure how similar two tasks are in their \textit{required action}, we
consider verb phrases within the task descriptions. The verb phrases in the task
description are chosen, as we expect them to reflect the actions that are
required to solve the task.
Two task descriptions that share some verb phrases are likely to be similar
regarding their \textit{required action}. However, many verb phrases bear
similar meaning, even though the vocabulary is not exactly the same. Therefore,
we apply word similarities from \textit{WordNet}
\cite{abdalgader_short-text_2010}, \cite{chang_integrating_2016}.
\subsubsection{Features for \textit{comprehensibility}}
To measure similarities in comprehensibility, we adopt features from the
structural and content feature set used in the classification approach. Those
features are: word count, number of bullet points, average words per sentence,
average commas per sentence, average chars per word, average paragraph length,
average line length, readability and lexical diversity.
Additionally, we add the \textit{ratio of unusual words} in the tasks'
descriptions, which is computed as the percentage of the words in the task's
description that are neither contained in more than five tasks in the whole
corpus nor in the English word list obtained from a Unix operating system.%
\subsubsection{Evaluation}
First experiments that apply the similarity measures to cluster the tasks in the
dataset show good results.
The category distribution of clusters regarding \textit{required action} (see
Table~\ref{table:clustering_analysis}) shows, that some clusters seem to model
known categories while others include tasks from many different categories.
\begin{table}[h]
	\begin{center}
		\caption[Category distribution in \textit{required action} clusters]{Category distribution for selected clusters regarding \textit{required action}.}
		\label{table:clustering_analysis}
		\begin{tabular}{llll}
			\toprule
			\textbf{Category} & \multicolumn{1}{l}{\textbf{A$_{1}$}} & \multicolumn{1}{l}{\textbf{A$_{6}$}}	& \multicolumn{1}{l}{\textbf{A$_{11}$}} \\
			\midrule
			Blog/Website Owners 	& - 	& - 	& 0.11 	\\
			Facebook 				& 0.08	& - 	& - 	\\
			Google 					& 0.10	& - 	& - 	\\
			Mobile Applications 	& 0.02	& - 	& 0.89 	\\
			Other 					& 0.16	& - 	& - 	\\
			Promotion 				& - 	& 0.03 	& - 	\\
			Search, Click, Engage 	& 0.47	& - 	& - 	\\
			Sign up 				& 0.09	& 0.94 	& - 	\\
			Youtube/Vimeo/... 		& 0.04	& - 	& - 	\\
			Various 				& 0.04	& 0.03 	& - 	\\
			\bottomrule
		\end{tabular}
	\end{center}
\end{table}
\section{Conclusion}\label{sec:conclusion}
This paper shows how the content of task descriptions can be used to create a
classification of micro-tasks. It also proposes two additional similarity
measures for micro-tasks. The evaluation shows that a classification is feasible
using the proposed setup. We also propose similarity measures that can be
applied to find similarities between micro-tasks. The proposed similarity
measures can model similarities, that are different from the known
categories.
%
%
\newpage
\bibliographystyle{aaai}
\bibliography{bibliography}

\end{document}